\documentclass[a4paper]{article}
\usepackage[letterpaper, portrait, margin=2cm]{geometry}
\usepackage[english]{babel}
\usepackage[utf8]{inputenc}
\usepackage{amsmath}
\usepackage{mathtools}
\usepackage{graphicx}
\usepackage{upgreek}
\usepackage[colorinlistoftodos]{todonotes}
\DeclareMathAlphabet{\mathcal}{OMS}{cmsy}{m}{n}
\SetMathAlphabet{\mathcal}{bold}{OMS}{cmsy}{b}{n}
\usepackage{braket}
\usepackage{url}
\usepackage{float}
\usepackage{array}
\usepackage{authblk}
\usepackage{graphicx, caption}
\usepackage{amsfonts}
\usepackage{braket}
\usepackage{amsmath,bm}

\usepackage[super,comma,square]{natbib}
\usepackage{amssymb}
\usepackage{amsthm}
\usepackage{authblk}
\usepackage{color}
\usepackage{mathtools}

\usepackage{bbm}
\usepackage{tikz}
\usepackage{setspace}

\newcommand{\maxE}{\mathop{\mathrm{max}}} 
\newcommand{\minE}{\mathop{\mathrm{min}}} 

\usepackage{abstract}

\usepackage{qcircuit}

\usepackage{blindtext}
\usepackage{multicol}
\usepackage[font=small,labelfont=bf]{caption}

\begin{document}

\title{Efficient Qubit Routing for a Globally Connected Trapped Ion Quantum Computer}
\author[1]{Mark Webber}
\author[2]{Steven Herbert}
\author[1]{Sebastian Weidt}
\author[1]{Winfried K. Hensinger}
\affil[1]{\small{\textit{Sussex Centre for Quantum Technologies, Department of Physics and Astronomy, University of Sussex,
Brighton, BN1 9QH, United Kingdom}}}
\affil[2]{\small{\textit{Department of Computer Science, University of Oxford, OX1 3QD, United Kingdom}}}
\date{2019}

\maketitle

\renewcommand{\abstractname}{}    
\renewcommand{\absnamepos}{empty} 

\begin{abstract} 
\noindent The cost of enabling connectivity in Noisy-Intermediate-Scale-Quantum devices is an important factor in determining computational power. We have created a qubit routing algorithm which enables efficient global connectivity in a previously proposed trapped ion quantum computing architecture. The routing algorithm was characterized by comparison against both a strict lower bound, and a positional swap based routing algorithm. We propose an error model which can be used to estimate the achievable circuit depth and quantum volume of the device as a function of experimental parameters. We use a new metric based on quantum volume, but with native two qubit gates, to assess the cost of connectivity relative to the upper bound of free, all to all connectivity. The metric was also used to assess a square grid superconducting device. We compare these two architectures and find that for the shuttling parameters used, the trapped ion design has a substantially lower cost associated with connectivity.

\end{abstract}

\begin{multicols}{2}

\section{Introduction}

Quantum computers are expected to solve classically intractable problems, such as accurately simulating the dynamics of large molecules \cite{Lanyon2010, Reiher2017}, which would greatly impact both material science and the pharmaceutical industry. In the finance industry even minor advantages can lead to significant returns \cite{Orus2019}. Phase estimation \cite{Kitaev1995} (for quantum chemistry) and Shor’s algorithm \cite{Shor} (for breaking RSA encryption), are two algorithms which promise an exponential speed up \cite{Nielsen2000, Montanaro2015}, but they both require a fault tolerant device for useful applications. Error correction techniques, such as the surface code \cite{Fowler2009,Fowler2012, Bermudez2017}, which facilitate a fault tolerant device have a very large physical to logical qubit overhead, requiring physical qubit numbers in the range of $10^{5}-10^{8}$. 

In recent years there has been growing interest and algorithmic development for Noisy Intermediate Scale Quantum (NISQ) computers \cite{Preskill2018} which do not require fault tolerance. The realization of ``quantum supremacy" \cite{Arute2019} represents a major milestone for such systems. Hybrid quantum algorithms such as the Variational Quantum Eigensolver \cite{Mcclean2016}, may provide an exponential speed up as compared to the classical counterparts. Assessing the capability of NISQ devices to run quantum algorithms is quite distinct from that of full-scale error-corrected quantum computers. For NISQ devices, this typically involves quantifying the achievable circuit depth of the device which represents the number of sequential gate operations that can be executed within the available coherence time. The analysis can be done in reverse to instead tailor near term algorithms to a specific device.

Superconducting circuits \cite{Wendin2017} and trapped ion devices \cite{Bruzewicz2019} are two of the leading quantum computing platforms. In particular, one architecture which offers a scalable approach to trapped ion quantum computing is based on a large connected ion trap array \cite{Kielpinski2002, Lekitsch2017}. This provides a solution to scale to very large qubit numbers, which will be a requirement to run many important algorithms. A key component of this architecture is the shuttling of individual ions to enable connectivity \cite{Kaufmann2017b,Kaushal2020,Hanneke2010}.

\begin{figure*}[t!]
\centering
\includegraphics[width=1\textwidth]{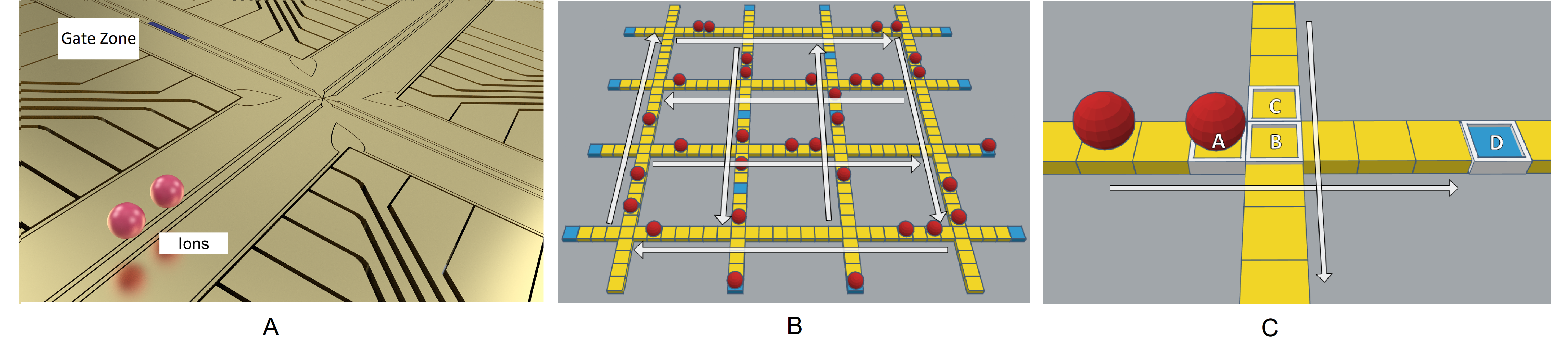}
\captionsetup{width=0.95\linewidth}
\caption{\label{fig:xjunction}\small{(A) A depiction of a single X junction which is repeated to form a grid on which the ions are restricted to, with zones dedicated to specific tasks. (B) A 3D representation of a quantum computing device using our proposed routing algorithm, where the yellow grid represents the X-Junctions, which the ions (red spheres) are restricted to, and the blue squares represent gate zones. The digitisation of the simulation can be seen with a resolution of 7 positions between adjacent X-Junctions. Arrows represent the lane priority of the routing algorithm. (C) A close up of an X-Junction from figure B. The routing logic used to decongest X-Junction centres involves occasionally ignoring the lane priority. Ions assigned to interior gate zones (blue square labelled D) have the closest X-Junction centre (labelled B) as their destination (one space off the centre because it is an area of lower trap stability (labelled A and C)). The ion in square A has been assigned to the local gate zone and it will travel back and forth between positions A and C directly, by ignoring the lane priority, to decongest for ions still travelling to their destination.}}
\end{figure*}

To utilize this shuttling based trapped ion architecture it is necessary to have a routing algorithm which can move large numbers of ions across the square grid array in parallel, and in an efficient manner. In this manuscript, we provide an ion routing algorithm which can enable arbitrary global connectivity, and we quantify its efficiency relative to a lower bound. When considering the achievable circuit depth of a NISQ device, one must include factors such as connectivity, gate fidelity, and the coherence time of the qubit. We provide an error model which can be used to estimate the achievable circuit depth of this quantum computing design as a function of experimental parameters.

Quantum computing architectures vary greatly, from the underlying system which represents the qubit, the available quantum gate set and to the means by which qubit connectivity is enabled. For superconducting architectures, qubits are stationary and connectivity is enabled through sequences of swap gates via nearest neighbor interactions, which will incur a high gate overhead for globally connected algorithms. For square grids with nearest neighbor connectivity, the best known method for globally connected algorithms on $N$ qubits scales with an overhead of $\Theta(N^{0.5})$ \cite{Cheung2007}, although it is only logarithmic if non-planar architectures are considered \cite{Herbert2018, Brierley2015} and optimisation of this swapping procedure is necessary to maximize performance \cite{Cowtan2019, Herbert2018a}. The characteristics of the desired algorithm will dictate the degree to which a device with inherent all to all connectivity outperforms a device which has a cost associated with enabling connectivity \cite{Linke2017}. The way in which connectivity is enabled varies greatly even within the trapped ion architectures. Architectures with stationary ions confined to a linear string benefit from global connectivity and multi-qubit gates \cite{Debnath2016, Schindler2013}, however as the number of ions co-existing in a single trap increases, it becomes progressively challenging to maintain key device specifications, such as gate fidelity. Furthermore, as the ion number $N$ increases, gate times increase as $\sqrt{N}$, and the increasing requirement on the number of motional modes will eventually lead to frequency crowding \cite{Brown2016}. Shuttling and swap operations may instead be used to enable connectivity by positioning multiple ions into the same region of space, where local gates may be performed. This introduces different challenges, but all required register reconfiguration operations have been demonstrated and several groups are further improving on aspects such as speed, and reliability. There are two main approaches to enable connectivity between trapped ion modules, one involves the use of photonic interconnects \cite{Monroe2012,Nigmatullin2016}, while the other, as described in the architecture analysed in this manuscript, utilises electric fields to connect adjacent modules. The connected modules form a continuous 2D plane, resulting in connection speeds between modules orders of magnitude higher as compared to photonic interconnects. 

The quantum computing architecture investigated in this manuscript consists of an ion trap array on a microchip, giving rise to a 2D grid to which all ions are restricted. The ions (where each ion represents a physical qubit) do not have to be stationary and are instead able to traverse the grid via shuttling operations. Entangling operations are performed by bringing the two (or more) desired ions to the same region of space (a gate zone). The smallest repeated unit of the architecture is the X-Junction (see Figure \ref{fig:xjunction}A). Logic gates may be performed by applying static voltages to a microchip in the presence of globally applied microwave fields and a local magnetic field gradient \cite{Weidt2016}. An alternative approach instead makes use of pairs of laser beams to execute quantum gates \cite{Fallek2016}, but this may be more challenging to implement for large numbers of qubits. This electronic microwave-based architecture has a clear path towards scaling to large qubit numbers \cite{Lekitsch2017}, and constitutes a practical blueprint for a quantum computer capable of solving some of the hardest problems, such as breaking RSA encryption. Furthermore, arbitrary two qubit connectivity can be enabled in near term devices relying only on ion shuttling operations (which can have a state fidelity comparable to stationary trapped ions \cite{Kaufmann2018}), without sequences of swap gates, as may be required in other architectures. To run an algorithm on a quantum computer based on this design, one first needs a routing algorithm which efficiently enables arbitrary connectivity between the ions in the square grid device, which is the main challenge addressed in this manuscript. The relevance of this manuscript is independent of the specific choice of gate operation, ion species and transition. Finding the optimum instruction set for each individual ion in real time is intractable and so we have solved the problem in a heuristic manner. The solution is motivated by one-way traffic flow with additional rule sets to deal with junction centres more efficiently. We quantify the efficiency of our approach relative to an unattainable lower bound and investigate its flexibility with regards to device shape, and ion density. We use these results, in combination with an error model we propose, to investigate the achievable depth and quantum volume for this design as a function of experimental parameters. We have made the error model publicly available \cite{Webber}. Quantum volume (QV) is a conceived metric for quantum computational power designed to enable sincere comparison between architectures \cite{Moll2017, Cross2019}, and we will discuss it in more detail in section 3.2.  

We have developed a simulation tool for the previously proposed architectural design of Lekitsch et al \cite{Lekitsch2017}. The simulation tool was used to develop and assess routing algorithms. The remainder of this manuscript is organized as follows: in section 2 we start by specifying the architecture and the connectivity problem to be solved, and then go on to explain the simulation tool and the developed routing algorithm. In section 3.1 we quantify the efficiency and versatility of our routing algorithm and in section 3.2 we present results on the achievable depth and quantum volume as a function of experimental parameters.

\section{Problem specification and \\ routing algorithm}

In the design being investigated here, ions (each encoding a single qubit) are restricted to a square grid (see Figure \ref{fig:xjunction}B) which consists of an array of repeated X-Junctions (see Figure \ref{fig:xjunction}A), each containing a single gate zone. Ions must first be shuttled (physically moved) into the gate zones for gates to be performed. The X-Junctions have a defined spacial resolution, which arises from the fixed number of electrodes on each arm but ions may be moved continuously. The gate zones enable both single and two qubit gates. To perform a quantum algorithm on this device it must first be decomposed into the native gate set, which can be optimized \cite{Maslov2017}. A decomposed quantum algorithm is defined by multiple rounds of gates, ideally all the required gates of an individual round will be applied in parallel, however the qubit number, gate density of the algorithm, and the number of gate zones will dictate the gate round overhead. In this architecture, each gate round is further broken into two parts, a routing sequence, where ions are shuttled into gate zones, which is then followed by the application of gates. We use the terminology ``shuttling" to refer to the act of moving ions in the device, and ``routing" to refer to the higher order logic of the shuttling. In this design, gates cannot be applied concurrently with shuttling. When the required number of gates in an individual round exceeds the number of available gate zones it is necessary to have multiple rounds of shuttling and gates, e.g. a gate round overhead of 2 would imply the need for: shuttle, apply gates, shuttle, apply gates. The shuttling round, which enables the connectivity, is the focus of this manuscript. When designing the routing algorithm, we optimized for the total time taken to enable global connectivity.

To enable arbitrary connectivity for this quantum computing architecture, we have created a simulation tool \cite{Webber} which represents the devices as a square grid consisting of iterated X-Junctions (see Figure \ref{fig:xjunction}B). The simulation is digitized to a variable resolution, where each position may contain 0, 1, or 2 ions. The ions are distributed evenly across the grid near the centre of each X-Junction and a quantum circuit (list of required two-qubit gates, i.e., ions that must be connected) is inputted. Ions which are assigned to the same gate zone are able to combine as a pair. Naive routing algorithms would not converge on a solution as ions with opposite travelling directions meet and cause permanent blockages. Positional swaps between ions have been demonstrated experimentally \cite{Kaufmann2017} and their usage would simplify the required routing algorithm. Here we present a solution that does not use swaps, and in section 3 we compare the effectiveness of routing both with and without swap operations. When bench-marking the device a randomly generated, globally connected, circuit was used. In order to assign ion (qubit) pairs to gate zones, we employ a greedy approach, assigning each pair to the nearest available gate zone (i.e., minimum combined distance of travel for the two ions), and addressing the pairs in an arbitrary order. This greedy approach is sufficient for a proof of principle using this prototype ion-routing algorithm, however we note that it may not yield the optimum gate-zone designations overall. To this end, a more sophisticated optimisation may be considered in future work, but we note that such combinatorial optimizations are generally hard problems themselves.

At each time step in the simulation, each ion is evaluated and moved sequentially according to the routing algorithm, which involves assessing its location, local environment and destination. The routing algorithm we have developed assigns alternating direction priorities to each lane of the square grid. The top-most horizontal lane is a right-only lane, the lane below it is left-only, and so on, and this also applies to vertical lanes (see Figure \ref{fig:xjunction}B). We ensure that the outer perimeter of the device is a clockwise loop regardless of the number of lanes, so that all gate zones can be reached, which means that odd size devices, e.g. one which consists of 3 by 3 X-Junctions, will not have fully alternating lane directions and instead will have, right, left, left, and up, down, down. We define a square grid device formed from M by M X-Junctions to be of device size M. We preferentially position gate zones on the exterior of the device where possible (on the outer arms of the perimeter X-junctions). Exterior gate zones are more favourable for routing as waiting ions do not block the movement of other ions. For square devices the number of interior gate zones scale with device size as $(M-2)^2$ and the exterior gate zones scale as $4M-4$, which results in a cross over point at device size 7 (98 qubits at 2 per X-Junction).

The centres of X-Junctions are decision points, where an ion will follow the lane priority towards its destination. Ions can enter the outer arms into the exterior gate zones only when it allows them to reach their assigned destination. Ions which are not destined to a gate zone during a given shuttling round have their destination set to their current location, and therefore only move to decongest. During development of the routing algorithm, a major bottleneck identified was congestion at interior gate zones. Devices larger than 2 by 2 have interior gate zones, and the ions waiting there can cause permanent blockages or unnecessary movement depending on how they are handled. To remedy this problem an additional feature was included, in which ions assigned to interior gate zones wait at the closest available X-Junction centre, where they are able to decongest efficiently by temporarily ignoring the lane priority (see Figure \ref{fig:xjunction}C). The movements available to each ion are dependent on multiple assigned parameters. The following binary questions determine these parameters: Does this ion have a destination for this round of gates? Is this a single ion or pair? Is this a waiting ion assigned to interior X-Junctions? The valid moves are then determined by using these parameters in combination with the location and local environment of the ion. At any particular time step an ion may have multiple valid moves available to it, hence there is a hierarchical list as follows from first priority to last: combine as a valid pair, a lane priority ignoring move as a waiting ion, a lane priority following move. Ions with no valid moves available will wait until the next time step.

\section{Results}
In section 3.1 we assess the efficiency and versatility of our routing algorithm. In section 3.2 we present an error model which utilizes the routing algorithm and includes experimental parameters such as gate fidelities, coherence time, ion loss and shuttling speed. We then use this error model to estimate the achievable depth and quantum volume of quantum computers based on this architecture. 

\begin{figure*}[t!]
\centering
\includegraphics[width=1\textwidth]{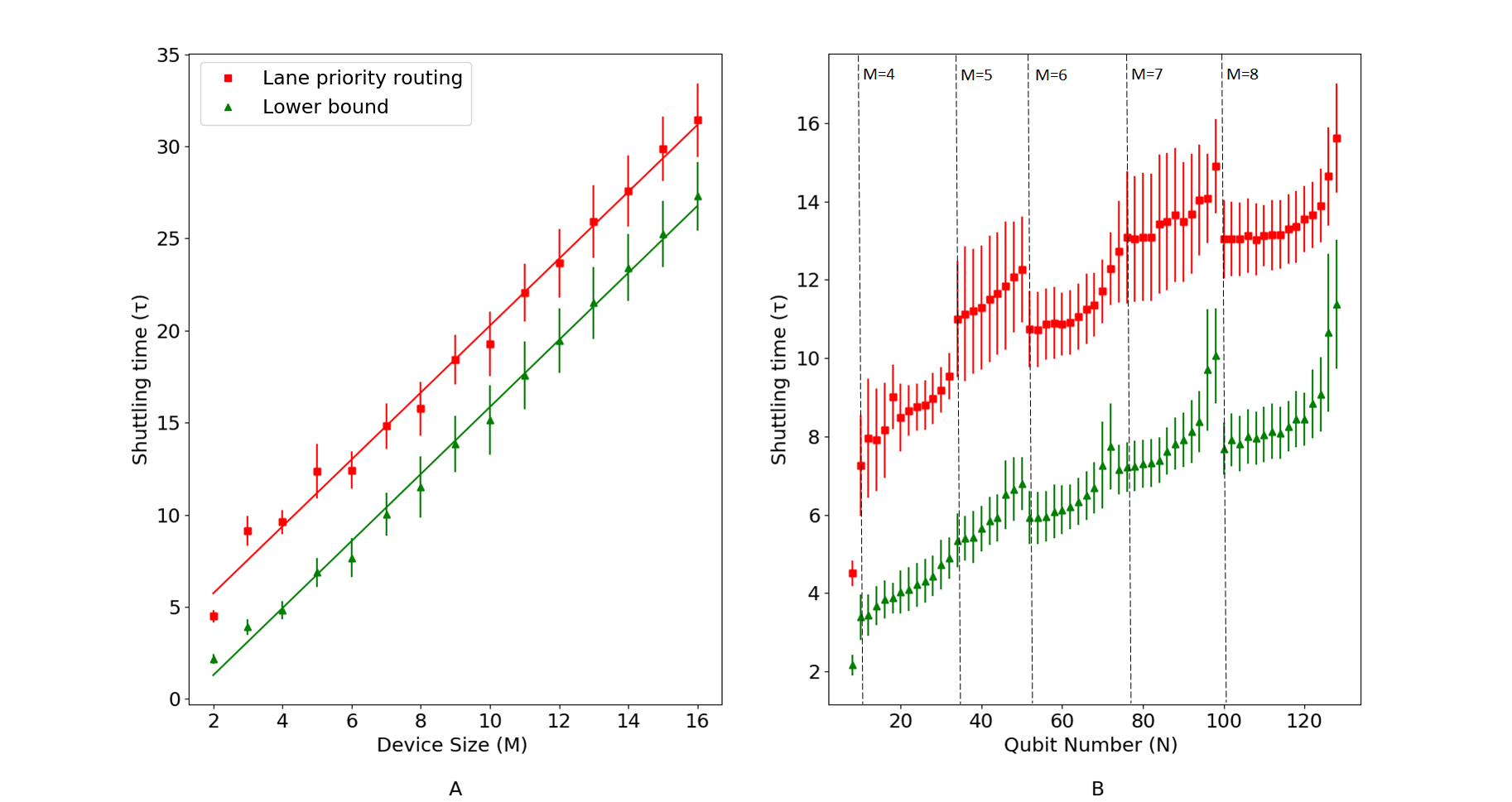}
\captionsetup{width=0.95\linewidth}
\caption{\label{fig:Shuttlescaling}\small{(A) Shuttling time, $\uptau$ (scaled by the resolution of the model so that a time of 1 is equal to the time it takes to shuttle between two adjacent X-Junctions), taken to enable connectivity as a function of device size (defined as a square grid consisting of M by M X-Junctions where M is the device size). There are two qubits initialized per X-Junction (plotted here for a range of 8 to 512 qubits). Red squares: Shuttling time for the routing algorithm. Green triangles: The lower bound (shortest route) shuttling time. The trend lines were generated using linear regression analysis and they both have a gradient of 1.82. Vertical lines represent one standard deviation. The results are the average value over 300 iterations of randomly selected pairings. The iteration number was chosen after investigating the mean and standard deviation convergence rate. (B) Shuttling time as a function of qubit number. The device size increases with qubit number when the device can no longer accommodate two qubits per X-Junction. Red squares: Shuttling time for our routing algorithm. Green triangles: The lower bound (shortest route) shuttling time. Vertical lines represent one standard deviation and the dashed lines mark where the device size is increased to accommodate the additional qubits}}
\end{figure*}

\subsection{Assessing the routing algorithm}

In this section we characterize the performance and flexibility of our routing algorithm, which we refer to as lane priority routing, within the framework of our abstract simulation tool. In section 3.2 we will introduce more practical considerations, allowing us to quantify the expected fidelity associated with enabling connectivity.  Randomly generated depth 1 circuits on $N$ qubits consisting of $N/2$ two qubit gates were iterated sufficiently to represent the requirement of global connectivity. After each iteration we count the total number of time steps which were required ($\uptau$), which can be converted into a total time $(s)$ by considering, the estimated speed at which one can shuttle between adjacent X-Junctions. It is important to note that increasing the speed of an individual shuttling operation may not always lead to an increase in the final fidelity, as the quality of the shuttling operation may impact on subsequent operations. At each iteration, a lower bound is calculated for that particular set of pairings, which is equal to the minimum number of time steps that will enable connectivity. To calculate the lower bound it is assumed that, qubits (ions) take the shortest path towards their destination and swap with no time penalty (i.e. the time required for an ion to move one discrete step is independent of whether a swap is performed or not). For a particular iteration, the ion with the greatest distance to travel is identified, and the number of spacial steps between its starting location and its destination is equal to the lower bound.

\begin{figure*}[t!]
\centering
\includegraphics[width=1\textwidth]{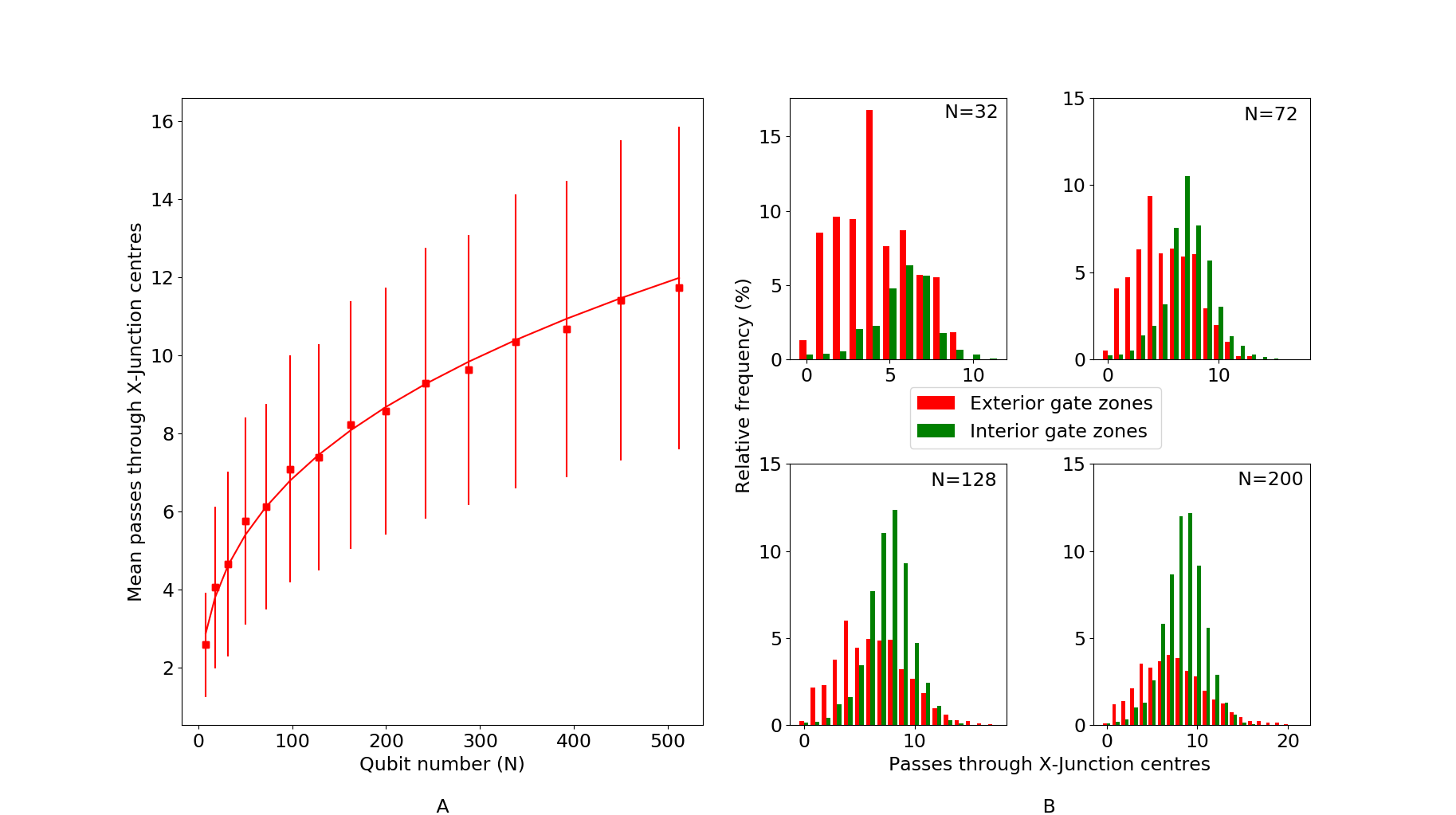}
\captionsetup{width=0.95\linewidth}
\caption{\label{fig:Xjunctionscaling}\small{(A) The mean number of passes through an X-Junction centre per ion as a function of qubit number for square devices with two qubits per X-Junction. Vertical lines represent a single standard deviation. (B) The relative frequency distribution of passes through X-Junction centres for four different device sizes, 4x4 (N=32), 6x6 (N=72), 8x8 (N=128), 10x10 (N=200). Red bars: Qubits assigned to exterior gate zones. Green bars: Qubits assigned to interior gate zones. 300 iterations of the globally connected depth-1 algorithm were used to generate a representative sample, and the frequency is scaled accordingly.}}
\end{figure*}

\begin{figure*}[t]
\centering
\includegraphics[width=1\textwidth]{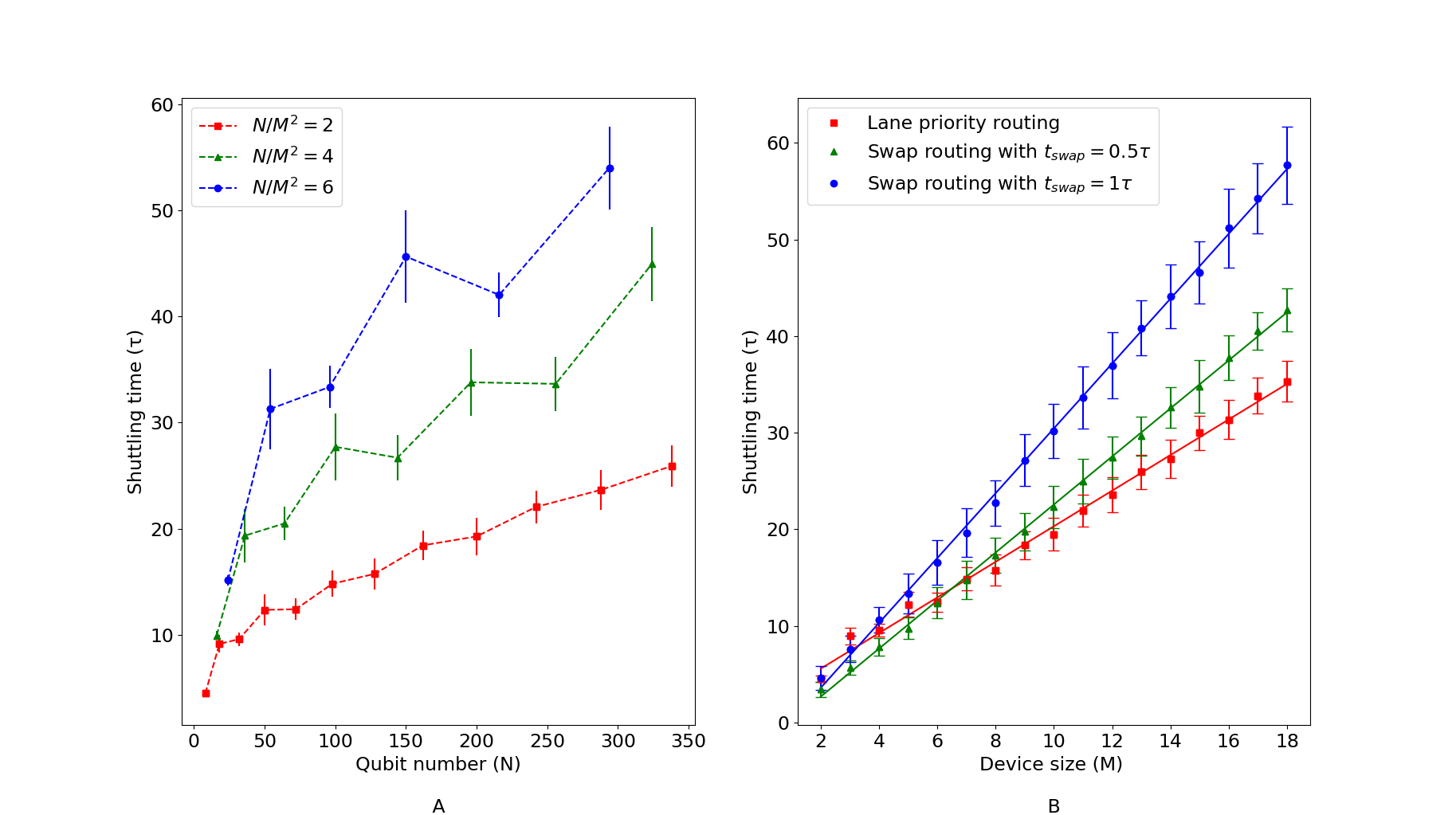}
\captionsetup{width=0.95\linewidth}
\caption{\label{fig:qubitdensities}\small{(A) Required shuttling time as a function of qubit number for three different qubit densities with red squares: 2 qubits per X-Junction. Green triangles: 4 qubits per X-Junction. Blue circles: 6 qubits per X-Junction. Vertical lines represent one standard deviation (B) Required shuttling time as a function of device size (defined as a square grid consisting of M by M X-Junctions where M is device size), with 2 ions per X-Junction, comparing swap based routing to our lane priority routing algorithm. Red squares: Lane priority routing. Green Triangles: Swap based routing with a swap time penalty equivalent to half the time it takes to shuttle between two adjacent X-Junctions. Blue Circles: Swap based routing with a swap time penalty equivalent to the time it takes to shuttle between two adjacent X-Junctions. Vertical lines represent one standard deviation and trend lines are fit using linear regression.  }}
\end{figure*}

The average shuttling time required to enable the global connectivity can then be compared to the lower bound as shown in figure \ref{fig:Shuttlescaling} A. These results are for devices with perfect two qubit gate parallelizability, i.e. there are two qubits initialized per X Junction. We conjecture that the total shuttling time would at best scale linearly with device size, $M$, because randomly selected distances in a square scales linearly with the length of the square. Both our routing procedure and the lower bound scale linearly with the device size and with a gradient of 1.82. There is a constant overhead associated with our routing which becomes less significant the larger the device is. The scaling for total shuttling time, $\uptau$, as a function of qubits, $N$, where $N=2\times M^{2}$ is $\uptau=1.3\,(3)\,\sqrt{N} +2\,(5)$, the fit and standard error were calculated using linear regression. An oscillating pattern on the lane priority routing results is noticeable with its relative magnitude decreasing with device size, which results from even sized devices outperforming odd sized devices. Odd sized devices (for example a device of 3 by 3 X-Junctions) cannot fully realize the alternating lane priority because we ensure that the outer perimeter lane is always a clockwise path. 

The routing algorithm is flexible and works well for a wide range of qubit numbers for a given device size. Figure \ref{fig:Shuttlescaling} B shows the shuttling dependence on qubit number for qubit densities less than or equal to 2 per X-Junction, i.e. with full gate parallelizability. The oscillating pattern resulting from odd and even device sizes is more notable. Shuttling time increases for both the lane priority routing and lower bound as more qubits are added to a device of static size, and peaks at a density of two qubits per X-Junction.  

The main criteria we optimized for when creating the routing algorithm was the total time. To calculate the achievable circuit depth at which a device can run,  the total error will not just be a function of the total time, but also include factors such as gate fidelity and ion loss. Traversing an X-Junction will have a corresponding ion loss rate which may be higher than the loss associated with linear shuttling. In order to quantify the associated error we have used our simulation to count the number of times qubits are expected to move through an X-Junction centre. The implications of these results for achievable depth will be explored in the following section. In figure \ref{fig:Xjunctionscaling} A the mean number of passes through an X-Junction, $X_{count}$, is plotted as a function of qubit number with vertical lines corresponding to a single standard deviation, and the dependence is well described by the following equation, $X_{count}=0.4\,(1)\,\sqrt{N} +2\,(2)$.  The distribution of passes is investigated in \ref{fig:Xjunctionscaling} B for four different device sizes, 4, 6, 8 and 10. The qubits are separated into two data sets, according to whether they are assigned to an interior or exterior gate zone. Across all device sizes investigated the maximum passes did not exceed 4x the stated mean. For the device with 72 qubits investigated in figure \ref{fig:Xjunctionscaling} B, the probability of an individual ion passing through an X-Junction centre $\geqslant$14 times is low, at approximately 0.2\%.

It may be desirable to increase the qubit density beyond 2 per X-Junction despite the potential loss of gate parallelizability as additional X-Junctions are experimentally costly to implement. Figure \ref{fig:qubitdensities} A shows the efficiency of the routing protocol for three different qubit densities. The increase in shuttling time is predominantly attributed to the multiple rounds of shuttling (and gates) which are required for the 100\% gate density (where gate density is the percentage of qubits involved in gates per time step) algorithm which we are assessing against. With a density of four qubits per X-Junction, a 100\% two qubit gate density algorithm would be completed by two full rounds of shuttling and gate applications. The oscillating pattern attributed to odd and even devices becomes more apparent with increasing qubit density. This analysis only includes the additional time associated with the multiple rounds of shuttling and does not include the gate time. The overall cost of increasing qubit density will depend on the gate density of the desired algorithm.

We created a new routing algorithm which relies on positional swaps where qubits take the shortest available route (ignoring the previously mentioned lane priority routing) and swap to decongest. We have compared the total shuttling time of the swap routing against the lane priority routing, for two different swap time penalties, shown in figure \ref{fig:qubitdensities} B. The time penalties were chosen based on early experimental results, H, Kaufmann \textit{et al} demonstrated fast ion swapping of 42$\mu s$ at a process fidelity of 99.5$\%$ \cite{Kaufmann2017}. Van Mourik \textit{et al} demonstrated positional ion swapping with an associated coherence loss of $0.2(2)\%$ \cite{VanMourik2020}. For ion shuttling speed, Walther \textit{et al} demonstrated fast shuttling of cold ions, over a distance of 280$\mu m$ in 3.6$\mu s$ \cite{Walther2012} and P, Kaufmann \textit{et al} demonstrated a state fidelity of 99.9994\%, for shuttling over a distance of 280$\mu m$ in 12.8$\mu s$ \cite{Kaufmann2018}. We characterized the average number of swaps, $n_{swap}$, per qubit for each connectivity run and found that for 18 qubits, the average was 1 swap, and for 50 qubits the average was 1.7 swaps. The dependence was well described by the following equation, $n_{swap}=0.23(2)\,\sqrt{N}+0.1(2)$, where the fit and standard error were calculated using linear regression. The average number of swaps per ion which was required to enable connectivity was found to be only weakly dependent on the swap time cost penalty, therefore doubling the time penalty results in minimal change to the number of swaps. For a wide range of device sizes the lane priority routing outperforms the swap based routing on total time taken, for the swap time penalties used here. This analysis suggests that for efficient routing in this 2D trap design, it will not be necessary to perform positional swap operations. However some combination of the lane priority routing and positional swaps may be favorable, depending on the expected costs associated with these operations. In the following section we will bring in more practical considerations to define the expected fidelity of a globally connected algorithm using our lane priority routing algorithm.

\subsection{Achievable depth and quantum volume estimations}

\begin{figure*}[t!]
\centering
\includegraphics[width=1\textwidth]{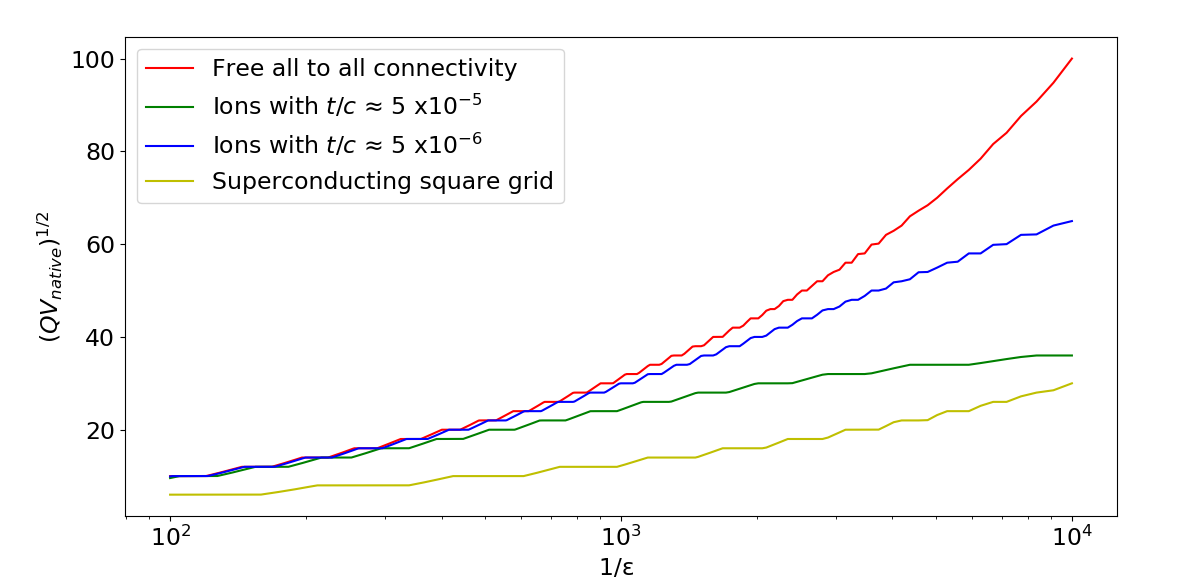}
\captionsetup{width=0.95\linewidth}
\caption{\label{fig:depthscaling}\small{Square root of the Quantum volume with a native two qubit gate requirement as a function of inverse gate error, $1/\epsilon$, for different architectures. Here, the number of qubits utilised to achieve a given value of $QV_{native}$ is equal to $(QV_{native})^{1/2}$ rounded up to the nearest even integer. Red: An architecture with all to all connectivity where $QV_{native}$ is solely defined by the native two qubit gate fidelity using equation 1 and represents the upper bound. Blue: The trapped ion architecture investigated in this manuscript using our proposed error model and the the routing results of the previous section. The coherence time and the time taken to shuttle between adjacent X-Junctions is extrapolated from work by Kaufmann \textit{et al} \cite{Kaufmann2018}. We assume a distance between adjacent X-Junctions of 2500$\mu$m \cite{Lekitsch2017} which implies a shuttling time, $t$, of 114$\mu$s, and we use the demonstrated state fidelity of shuttling \cite{Kaufmann2018} to infer a coherence time, $c$, of 2.13s, and so $t/c	\approx 5 \times 10^{-5}$. We assume an ion loss rate of $10^{-5}$ per X-Junction pass. We assume each iteration of the depth-1 circuit requires one combination and one separation operation, each of which have a duration of 80$\mu$s \cite{Ruster2014, Bermudez2017}, and we assume the state fidelity of the operation can be inferred from the coherence time. Green: All the assumptions are identical to the above except for the coherence time which has been increased by a factor 10 \cite{Harty2014}. Yellow: A model of a square grid superconducting architecture where connectivity is enabled through sequences of nearest neighbour swap operations which require 3 native two qubit gates (the CNOT). The depth overhead was found to scale as a function of qubit number N as $2.77\sqrt{N}-4.53$ using the publicly available quantum compiling software, CQC's $t\ket{ket}$; improvements to the connectivity compiler would reduce this overhead.
}}
\end{figure*}

For comparison between near term quantum computers, one must consider more than just the number of qubits. Quantum volume (QV) is a conceived metric for quantum computational power designed to enable sincere comparison between architectures \cite{Moll2017, Cross2019}. QV includes factors such as gate fidelity, qubit number, connectivity and the available gate set. Below we use the definition as given by Moll et al \cite{Moll2017} (which has some differences to the definition given by Cross et al \cite{Cross2019}).

\begin{equation}
\label{eq10}
	QV =\maxE_N \Big[ \minE \Big[ N, \frac{1}{N \times \epsilon_{eff}(N)}\Big]^2\Big],
\end{equation}

for the number of qubits within the device $N$, and effective error rate $\epsilon_{eff}$, which typically depends on $N$. QV reflects the limiting factor of the device, which is either the qubit number or the achievable depth $D$, where $D=1/(N \times \epsilon_{eff}(N) )$. To compute QV, a randomly generated depth-1 circuit on $N$ qubits with $N/2$ arbitrary (SU(4)) two qubit gates is used. The achievable depth represents the circuit depth at which the device can run before coherence is lost, specifically, the depth at which at least one qubit error is statistically likely. The achievable depth is a useful metric which can be used separately from QV to estimate the feasibility of running an algorithm on a NISQ device. 

The effective error $\epsilon_{eff}$ for each depth-1 circuit includes gate error, and errors associated with gate decomposition, connectivity and parallelizability. The effective error can be used to calculate the achievable depth. Many iterations of the randomly generated circuit should be used to best capture the properties of the device. In this section we present an error model for the quantum computing design analysed in this manuscript and present results for a range of experimental parameters that may be achievable. 
In the following analysis we assume linear propagation of errors, which represents a worst-case outcome, as it does not account for the possibility of a new error reducing a previous error. We combine the errors associated with connectivity and gates, as opposed to a full simulation of the quantum states and associated noise model. The advantage of this methodology is that we are able to make estimations on effective error (and therefore achievable depth) for a wide range of qubit numbers and device sizes. The effective error $\epsilon_{eff}$ for this design and circuit requirement is, $\epsilon_{eff}=\epsilon_{gate}+\epsilon_{conn}$, where $\epsilon_{gate}$ is the two qubit gate error and $\epsilon_{conn}$ is the error associated with enabling the required global connectivity. We decompose $\epsilon_{conn}$ into two components $\epsilon_{conn}=\epsilon_{deco}+\epsilon_{loss}$ where $\epsilon_{deco}$ is the quantum decoherence associated with the total time taken to enable connectivity where $\epsilon_{deco} = 1-e^{-t/c}$ for time $t$ and coherence time $c$. Recent work by Kaufmann \textit{et al} \cite{Kaufmann2018} demonstrated high state fidelity shuttling (99.9994\%), where the coherence time associated with shuttling was extrapolated to be $2.13s$. A coherence time of $50s$ has been demonstrated for stationary ions in the atomic clock states of calcium \cite{Harty2014}. In figure \ref{fig:Shuttlescaling} A we quantify the average time required to enable connectivity as a function of device size (qubit number). The stated dimensionless time $\uptau$ can be converted to a real time by multiplying it with the expected time to shuttle an ion between two adjacent X-Junctions. For ion shuttling speed, a distance of 280$\mu m$ has been demonstrated in 3.6$\mu s$ \cite{Walther2012} and 12.8$\mu s$ \cite{Kaufmann2018}. There will be an additional time cost associated with performing a single combination and a separation of ions, per iteration of the depth-1 circuit, which have been performed in 80$\mu$s \cite{Ruster2014, Bermudez2017}. $\epsilon_{loss}$ represents the likelihood for an ion to be lost to the vacuum per iteration of the depth-1 circuit. Investigations of ion loss for routing across X-Junction centres \cite{Wright2013} found continuously Doppler cooled ion survive more than $10^5$ round trips whereas uncooled ions survive at least $65$ round trips. Ion loss occurs when its motional energy exceeds the trap depth, which can be remedied by increasing the trapping potentials and by cooling techniques. Significant work is carried out in order to allow the application of large trapping voltages in order to increase the effective trapping potential; recently trapping voltages as large as 1000V have been demonstrated \cite{Sterling2014}.  In figure \ref{fig:Xjunctionscaling} A we quantify the average number of X-Junction crosses, $X_{count}$, as a function of device size (qubit number), which can be combined with an ion loss per shuttle rate, $X_{loss}$, for $\epsilon_{loss}$. This can all be combined into a single equation defining the effective error in this design

\begin{equation}
\epsilon_{eff}=\epsilon_{gate}+(1-e^{-t/c})+ (X_{count} \times X_{loss})
\end{equation}

This error model can be used to estimate the achievable depth for a wide variety of device sizes and experimental parameters for devices following this design. The gate error will depend on the requirement of the algorithm we are assessing against, which in the case of QV is the arbitrary two qubit gate.  The focus of this manuscript is the cost of enabling connectivity, therefore we have chosen to utilise the concept behind QV but alter its algorithm requirement to instead be the native two qubit gate of the architecture being assessed. We will refer to this new metric as $QV_{native}$ going forward. The costs associated with arbitrary two qubit gate decompositions will be discussed later. 

We use our error model to quantify $QV_{native}$ as a function of two qubit gate fidelity for this architecture with two different assumptions on experimental shuttling parameters, shown in figure \ref{fig:depthscaling}. These can be compared to the upper bound of this metric which corresponds to a hypothetical architecture with free, all to all connectivity. To demonstrate an example, a device with free all to all connectivity and a two qubit gate fidelity of 99.9\% has a  $(QV_{native})^{1/2}$ of 31.25. This implies that one could effectively run a globally connected native two qubit gate algorithm with approximately 30 qubits at depth 30.  We investigate up to a two qubit gate fidelity of 99.99\%; this analysis indicates that without error correction techniques, chasing high qubit numbers will be futile even with considerable improvement to the current state of the art two qubit gate fidelities. The trapped ion plots of figure \ref{fig:depthscaling} have an ion loss rate of $10^{-5}$; we found that increasing this rate substantially decreases the $QV_{native}$, which seriously emphasises the importance of achieving an ion loss rate of this order.  The ion loss rate can be improved by deeper trapping potentials and by techniques such as sympathetic cooling. Ions may also be automatically reloaded from a filled reservoir trap section.

We also quantify this metric for a model of a superconducting architecture, which is a square grid with only nearest neighbour interactions. In superconducting square grid systems, connectivity is enabled by sequences of swap operations, and the best known method has an overhead of $\Theta(N^{0.5})$ \cite{Cheung2007} for the random complete graph (global connectivity). IBM provide an equation to estimate the depth overhead, of the form $ (a\sqrt{N} +b)$, for a square grid but it includes their gate decomposition costs of arbitrary two qubit gates\cite{Cross2019}. Cowtan \textit{et al} developed a compiler to map quantum circuits to devices with restricted qubit connectivity and provides results on the depth overhead for nearest neighbour square devices \cite{Cowtan2019}. Using the publicly available software, CQC's $t\ket{ket}$ and its recently improved connectivity compiler, the depth overhead was found to scale with qubit number $N$ as, $2.77 \sqrt{N} -4.53$. This overhead corresponds to a depth-1, 100\% gate density, native two qubit gate (CNOT) algorithm with $10 N$ iterations. A SWAP gate is implemented with three CNOTs and no advantageous initial qubit mapping was utilised.

The native two qubit gate of this trapped ion design is the Mølmer-Sørensen \cite{Molmer1998} and although it does not directly depend on the motional state, it is affected by the heating rate and experimental offsets whereby it is favourable to begin in a low motional state. Therefore to reach the high two qubit gate fidelities used in figure \ref{fig:depthscaling}, it will be necessary to use cooling techniques. Techniques such as Doppler and sideband cooling are only suitable for the beginning of a quantum algorithm as they do not preserve quantum information. Sympathetic cooling is a way of actively cooling throughout a quantum algorithm, whereby the qubit is sympathetically cooled via a different laser cooled ion species. It is likely to be a critical technique for the use of trapped ion devices, particularly in the fault tolerant regime. Shuttling based designs may benefit from multi-species shuttling. The relative difference between the upper bound of free all to all connectivity, and the plots for trapped ions, increases with the two qubit fidelity due to the independent cost associated with shuttling.   We find a notable difference in the $QV_{native}$ between the superconducting plot and the all to all, particularly at higher two qubit gate fidelities. Superconducting square grids have a slower growth rate with two qubit gate fidelity because the associated depth overhead of swaps increases with the number of qubits (the size of the grid). In this model, the trapped ion design outperforms the superconducting square grid for this set of experimental shuttling parameters. The number of shuttling operations, $\uptau$, required to enable connectivity in the trapped ion design analysed here, scales as $\uptau=1.3\,(3)\,\sqrt{N} +2\,(5)$ which is comparable to the depth overhead for swapping on the superconducting square grid. Extrapolating from the high state fidelity shuttling of Kaufmann \textit{et al} \cite{Kaufmann2018}, it implies a fidelity per shuttling operation (2500$\mu m$) of 99.995\% which is significantly higher than the two qubit gate fidelities achieved so far by superconducting systems. In order to facilitate further work with our error model by others, we have made it open access \cite{Webber}. To experimentally implement the work presented here, a key challenge is to build and operate such a trap as shown in Figure \ref{fig:xjunction}B. A trap needs to be fabricated for which a number of approaches are being perused \cite{Romaszko2019}.

The QV metric requires application of arbitrary, randomly generated SU(4) two qubit gates, as opposed to the native two qubit gate investigated above. The purpose of this requirement is to capture the power of the architecture's native gate set. There is an upper bound circuit which can express any arbitrary U(4) using 3 CNOTs and 15 elementary single qubit gates \cite{Vatan2004}, with a native gate set consisting of $R_{x} (\theta), R_{z} (\theta)$, and the CNOT. We have translated this upper bound circuit into the native gate set of the architecture analysed here, which is $R_{x} (\theta), R_{y} (\theta)$ and the Mølmer-Sørensen (MS) two qubit gate \cite{Molmer1998} (see Appendix). The gate count of the new upper bound circuit is 3 MS gates and 18 elementary single qubit gates. We reduced the initial single qubit gate count from 29 to 18 by utilising basic commutation relations and the degrees of freedom which are available \cite{Maslov2017}. The upper bound circuit represents a worst case and optimal circuits can be found for particular SU(4)s using analytical techniques \cite{Blaauboer2006a} but most exact decompositions of arbitrary two qubit gates will require the three native two qubit gates of the upper bound. A new technique demonstrated by IBM can considerably improve the fidelity of decomposing these gates \cite{Cross2019}; Cross \textit{et al} instead start with an allowable error on the decomposition, which allows one to identify cases which require less than the upper bound of three two qubit gates. This can result in a considerable improvement to the final fidelity, particularly when working with lower two qubit gate fidelities. The quantum volume with native qubit gates we have used in this section is a clear tool of comparison for the cost of connectivity in these two architectures. To extend this comparison to architectures with drastically different gate sets, such as those in some trapped ion designs which enable multiple ($>2$) qubit gates, the original QV metric is more suitable. Once more research characterising quantum volume for various quantum computing designs becomes available, a more detailed comparison would be warranted.

\section{Conclusion}
The quantum computing architecture analysed in this manuscript has a clear path towards scaling to large qubit numbers. Arbitrary connectivity between qubits can be enabled in this design on near term devices, relying only on shuttling across a square grid, but prior to this work there were no proposed routing algorithms. We have created a routing algorithm which efficiently enables connectivity in this design. A simulation tool was created which allowed us to characterize the routing algorithm and compare it against a strict lower bound to which it scales with an equal gradient. The routing algorithm compares favourably against positional-swap based routing for the experimental values used. We propose an error model which can be combined with the results from the simulation tool, to estimate the circuit depth of a device as a function of experimental parameters. We use a metric, $QV_{native}$, based on quantum volume which instead has native two qubit gates, to focus on and assess the cost of connectivity in this trapped ion design. The ion loss per shuttling operation was found to be an important parameter of the model and needed to be low, at $10^{-4}$ - $10^{-5}$, to reach appreciable circuit depths and it can be improved experimentally with larger trapping potentials. It is necessary to maintain a sufficiently low motional state energy of the ions to reach high two qubit gate fidelities, which highlights the importance of developing techniques such as sympathetic cooling, and therefore multi-species shuttling. We use $QV_{native}$ to assess a model of a square grid superconducting device, and find that for the shuttling parameters used, this trapped ion design has a substantially lower cost associated with connectivity. The simulation tool and this analysis can be used to inform the development of devices following this design, by metering experimental priorities, and by solidifying the requirements on shuttling. This work has implications for error correction schemes, especially those which rely on non-nearest neighbor interactions.

\section{Acknowledgement}
This work is supported by the U.K. Engineering and Physical Sciences Research Council via the EPSRC Hub in Quantum Computing and Simulation (EP/T001062/1),  the U.K. Quantum Technology hub for Networked Quantum Information Technologies (No. EP/M013243/1), the European Commission’s Horizon-2020 Flagship on Quantum Technologies Project No. 820314 (MicroQC), the U.S. Army Research Office under Contract No. W911NF-14-2-0106,  and the University of Sussex. Thanks are given to David Bretaud for helpful conversations regarding gate decomposition, to Zak Romaszko for help with 3D modelling, to Mitchell Peaks for proofreading, and finally to Silas Dilkes for help with utilising $t\ket{ket}$ to calculate the superconducting square grid swap cost.  

\bibliography{REFS} 
\bibliographystyle{ieeetr}

\end{multicols}

\section {Appendix}
\subsection{Decomposing arbitrary two qubit gates}
 
An upper bound circuit for expressing arbitrary two qubit gates in terms of $R_{x} (\theta), R_{z} (\theta)$, and the CNOT, was found by Vatan \textit{et al} \cite{Vatan2004}.

\[ \Qcircuit @C=1em @R=.7em {
     & \gate{A_1} & \targ & \gate{R_z} &\ctrl{1} & \qw & \targ & \gate{A_3} &\qw  \\
     & \gate{A_2} & \ctrl{-1} & \gate{R_y} & \targ & \gate{R_y}&  \ctrl{-1} & \gate{A_4} & \qw
}\]

\begin{center}{\small{\textbf{Figure 6:} A circuit for implementing any transform in U(4) with a gate set consisting of $R_{x} (\theta), R_{z} (\theta)$, and the CNOT, where the gate $A_i$ here represents an arbitrary single qubit transform, for a total gate count of 15 elementary single qubit gates and 3 CNOTs.}}\end{center}

\noindent
An arbitrary single qubit gate $U_1$, can be expressed in the form, 

\begin{equation}
U_1=e^{i\alpha}R_{\hat{n}}(\beta)R_{\hat{m}}(\gamma)R_{\hat{n}}(\delta)
\end{equation}
\noindent
for appropriate choices of $\alpha,\beta,\gamma,\sigma$, where $\hat{n}$, and $\hat{m}$ are non-parallel real unit vectors in three dimensions \cite{Nielsen2000}. We have converted the circuit of figure 6 into the native gate set of the architecture investigated here, which is, $R_{x} (\theta), R_{y} (\theta)$ and the Mølmer-Sørensen gate $U_{MS}(\chi)$ \cite{Molmer1998} which has the form,

\begin{equation}
U_{MS}(\chi) =
\begin{pmatrix}
cos(\chi)&0&0&-isin(\chi)\\
0&cos(\chi)&-isin(\chi)&0\\
0&-isin(\chi)&cos(\chi)&0\\
-isin(\chi)&0&0&cos(\chi)\\
\end{pmatrix}
\end{equation}
\\
\noindent
where $\chi$ can be set between $-\pi/4$ and $\pi/4$. The new converted circuit is shown in figure 7, and has a gate count of 3 MS gates and 18 single qubit gates. The single qubit gate count was reduced by combining superfluous sequences of single qubit gates, utilising commutation relations, and the available degrees of freedom. The MS gate commutes with any $R_x(\theta)$. When decomposing the CNOT and $R_z(\theta)$ gate, there is an available degree of freedom, where one may choose the direction of rotation on certain $R_y$ gates, which can then be used to eliminate some $R_y$ gates from the circuit \cite{Maslov2017}.

\[ \Qcircuit @C=1em @R=.7em {
     & \gate{R_y R_x R_y}& \multigate{1}{MS} & \gate{R_y R_x} & \multigate{1}{MS} & \gate{R_y} & \multigate{1}{MS} &  \gate{R_y R_x R_y} &\qw  \\
     & \gate{{R_y R_x R_y}}  & \ghost{MS} & \gate{R_y R_x} & \ghost{MS} & \gate{R_y} &  \ghost{MS}& \gate{{R_y R_x R_y}} & \qw
}\]

\centering \small{\textbf{Figure 7:} A circuit for implementing any transform in U(4) with a gate set consisting of $R_{x} (\theta), R_{y} (\theta)$, and the Mølmer–Sørensen gate \cite{Vatan2004}, for a total gate count of 18 elementary single qubit gates and 3 MS gates.}

\end{document}